\documentclass[12pt]{iopart}
\expandafter\let\csname equation*\endcsname=\relax 
\expandafter\let\csname endequation*\endcsname=\relax 
\usepackage{amsmath,amssymb,amsthm}

\usepackage{iopams}
\usepackage{graphicx}
\usepackage{subcaption}
\usepackage{hyperref}
\usepackage{physics}
\usepackage{jabbrv}
\usepackage{latexsym}
\usepackage{xcolor,soul}
\usepackage[normalem]{ulem}

\begin{document}
\title[Coherent detection schemes for SCW CV-QKD]{%
  Coherent detection schemes for subcarrier wave continuous variable
  quantum key distribution
}

\author{E~Samsonov$^{1,2}$, R~Goncharov$^1$, M~Fadeev$^1$, A~Zinoviev$^1$, D~Kirichenko$^1$, B~Nasedkin$^1$, A~Kiselev$^1$, and V~Egorov$^{1,2,3}$}

\address{$^1$ ITMO University, Kronverkskiy, 49, St.Petersburg, 197101, Russia\\
$^2$ Quanttelecom LLC., Saint Petersburg, 199178 6 Line 59, Russia\\
$^3$
National Center for Quantum Internet, Kronverkskiy, 49, St.Petersburg, 197101, Russia}

\ead{eosamsonov@itmo.ru}

\begin{abstract}
We examine different methods to implement coherent
detection in
the subcarrier wave quantum key distribution (SCW QKD)
systems.
For classical wavefields,
we present the models
describing homodyne-type
and heterodyne-type coherent detection schemes
needed to
extract information from the quadrature phase-coded
multimode signals
used in SCW QKD.
Practical feasibility of the
proposed schemes is corroborated by the experiments.
\end{abstract}

\vspace{2pc}

\submitto{\QST}

%%%%%%%%%%%%%%%%%%%%%%%%%%  body  %%%%%%%%%%%%%%%%%%%%%%%%%%
\section{Introduction}
\label{intro}

Since 1896, when the "beat receptor", the first heterodyne detector,
as we know it now, was invented by Nicola Tesla~\cite{tesla}, various
forms of coherent detection methods, including homodyne and heterodyne
detection, have found wide use in a variety of applications.
Starting with radio
communications~\cite{roger1952double},
the coherent detection methods
have been expanded to the optical domain~\cite{protopopov}.
Nowadays there
are numerous examples of optical coherent detection usage
in different fields including telecommunications 
% ~\cite{smith1990spread, strand2006compact,shimizu1993coherent,dong2013generation}
~\cite{ly2006coherent,tanosaki2003vivo,kikuchi2015fundamentals,mecozzi2018coherent,lu2010distributed}
and quantum
optics~\cite{yuen1983noise,barchielli1990direct,wallentowitz1996unbalanced}.
One
of the promising technology at the boundary between these is
the quantum
key distribution (QKD), which, in combination with coherent detection,
leads to a separate branch, known as the continuous variable (CV)
QKD~\cite{grosshans2003quantum,hirano2003quantum,leverrier2011continuous,heid2006efficiency,
  bradler2018security,
  papanastasiou2018quantum,ghorai2019asymptotic,lin2019asymptotic,zhang2019integrated,zhang2020long}.
The CV-QKD
systems rely on the methods of coherent detection for gaining information
encoded in the electromagnetic field quadratures. In other words,
single-photon detection can be replaced by conventional optical
communication methods. 

Active development of novel QKD systems utilising
modulated multimode
states of light
necessitates
 reconsideration and modification
 of the existing
detection schemes.
A striking example of such a system
is
the subcarrier wave (SCW)
QKD~\cite{merolla1999phase,mora2012experimental,gleim2016secure,miroshnichenko2018security,gaidash2019methods,chistiakov2019feasibility,kynev2017free,samsonov2019continuous}. 
The method for quantum state encoding
is the distinguishing feature of the SCW QKD.
In this method, a strong
monochromatic wave emitted by a laser is modulated in an
electro-optic phase modulator to produce weak sidebands,
whose relative phase with respect to the strong (carrier) wave
encodes the quantum information
(a detailed description of the discrete variable (DV) SCW QKD
conventional protocol
and the SCW CV-QKD
protocol
can be found in~\cite{gleim2016secure}
and~\cite{samsonov2019continuous}, respectively). 
% SCW QKD protocols implemented to date are discrete, so the weak
% radiation component is detected by a single photon counter, and the
% measured observable has a discrete spectrum.

In this paper we present three coherent detection schemes with phase
estimation to demodulate the quadrature phase-coded multimode
signals. This type of signals is used to encode quantum information in
the SCW QKD systems.
The main advantage of the proposed schemes is using the
carrier wave which plays a leading part in
the SCW methodology as a local
oscillator (LO).
In practice, it provides an alternative solution of the well-known problem of
transmitting the local oscillator through the quantum channel (or its
generation on the receiver’s side).
This is the novel approach that has not
been previously discussed in works on the multimode
CV-QKD~\cite{fang2014multichannel,gyongyosi2019subcarrier,wang2019optical}.
So we present the theoretical models describing the proposed detection schemes and
experimentally demonstrate how to implement coherent
detection using the carrier wave as a local oscillator.

\section{Phase-coded multimode signals}
\label{sec:phase-coded}
Our first step is to describe generation of the phase-coded multimode
signals that is the essential part of the SCW QKD system.
Since
our proof-of-principle experiments use classical light,
we shall utilise the classical model of electro-optic phase
modulation~\cite{haykin2008communication}.
Note that quantum description of
electro-optic modulation applicable to quantum states can be found
in~\cite{miroshnichenko2017algebraic,capmany2010quantum,kumar2008evolution}.

\begin{figure}[ht]
\centering
\includegraphics[width=0.7\textwidth]{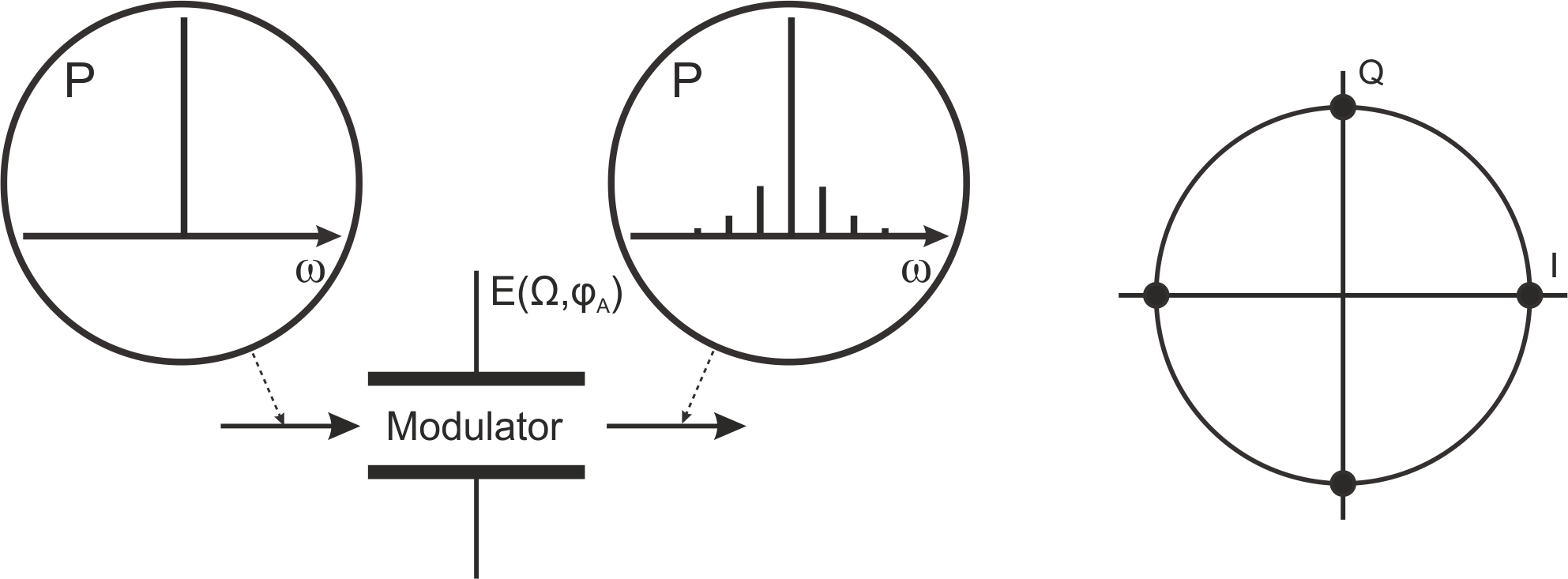}
\caption{Modulated light emerging after the electro-optic phase
  modulator and 4-PSK constellation diagram. The typical power
  spectrum at the modulator output in SCW QKD is also shown.}  
\label{fig1}
\end{figure}

As is shown in Figure~\ref{fig1},
a monochromatic lightwave  with the frequency
$\omega$ and the amplitude $E_0$ is modulated by
the traveling wave electro-optic phase modulator using
the microwave field with
the frequency $\Omega$ and the phase
$\varphi_{a}$~\cite{yariv1984optical}.
The phase-modulated optical
field $E(t)$ then can be expressed in terms of Jacobi–Anger expansion as
follows 
\begin{equation}
\label{e1}
E(t) =  E_0e^{i\omega t}e^{i m_a \cos{(\Omega t+\varphi_{a})}}=
E_0e^{i\omega t}
\sum_{k=-\infty}^{\infty}{i^kJ_k(m_a)e^{ik(\Omega t+\varphi_{a})}},
\end{equation}
where  $J_k(m_a)$ is the $k$-order Bessel function of the first kind
and $m_a$ is the modulation index
which is typically small in the SCW QKD protocols.
As a result, the sidebands are formed at
the frequencies $\omega_{k}=\omega+k\Omega$,
where $k$ is the integer.  

The modulated light beam emerging at the output of the phase modulator
contains the reference carrier ($k=0$) and the sidebands ($k\neq 0$). The
relative phase and amplitude between reference and all the sidebands
are determined by the inner microwave field that controls the
phase $\varphi_{a}$ and the modulation index $m_a$, respectively.
Therefore
one may encode the information into the sidebands of the phase-modulated light
by applying different forms of quadrature amplitude
modulation~\cite{mora2012experimental,gleim2016secure,gonzalez2010optical,gasulla2012phase,zhang2008photonic}.
In this paper, for simplicity,
we concentrate on the quadrature phase-shift
keying which is extensively used in CV-QKDs
with discrete
modulation~\cite{lin2019asymptotic,ghorai2019asymptotic,hirano2017implementation,
  leverrier2009theoretical}.
We prepare the phase-coded multimode signals
by selecting the phase of the microwave field from the finite set of
$\varphi_a\in\{0,\:\pi/2,\:\pi,\:3\pi/2\}$
which is equivalent to the four states commonly used in
the CV QKD protocols with discrete
modulation.
An example of the modulated signal constellation diagram
is presented in
Figure~\ref{fig1}.

%%%%%%%%%%%%%%%%%%%%%%%%%%%%%%%%%%%
\section{Subcarrier wave implementations of coherent detection}
\label{sec:scw-implem}
%%%%%%%%%%%%%%%%%%%%%%%%%%%%%%%%%%%%

%%%%%%%%%%%%%%%%%%%%%%%%%%%
\subsection{Classical coherent detection}
\label{subsec:class-detection}
%%%%%%%%%%%%%%%%%%%%%%%%%%%%%%%%

For comparison purposes,
we begin with a brief discussion of
the basic concepts of coherent detection which is based on
using a $50/50$ beam splitter to
mix
an initial weak signal of the power $P_{s}$ and the reference field
generated by an external source
of the power $P_{LO}$
also known as the local oscillator (LO).
The mixed optical fields at the outputs of the beam splitter
are detected by a photodiode,
and the difference between the signals
is registered  by a balanced detector.
It turned out that the output signal which power is
proportional to the square root of LO power, $\sqrt{P_{LO}}$,
carries the information about the initial signal.
In addition, subtraction of the signals
results in reduction of noise components,
whereas
the amplification scheme installed in
the balanced detector is used for
further amplification of its response.

%Homodyne and heterodyne detection methods are widely used in modern
%fiber-optic communication systems due to the possibility of
%implementing various types of multilevel modulation. In the case of
%QKD systems, it makes it possible to register a weak signal without
%using bulky single-photon detectors, which is the main advantage over
%DV QKD systems. 

In the case of homodyne detection, the signal frequency $\omega_{s}$ and
the LO frequency
$\omega_{LO}$  are equal.
The constant signal level
that will be observed
is determined by
the phase difference between the initial signal
and the LO,
$\Delta\varphi=\varphi_{s}-\varphi_{LO}$.
For instance,
when the constant signal on the oscilloscope is positive
at zero phase difference,
it will be negative provided the
phase difference equals $\pi$.
More precisely,
in the absence of noise,
the output of the
detector is proportional to the difference between the
photocurrents registered by the photodiodes
and,
in the absence of noise,
can be written in the following form~\cite{kikuchi2015fundamentals}: 
\begin{align}
  \label{currentref}
  &
    I(t) =  I_1(t)-I_2(t)=2R(\lambda)G\sqrt{P_s(t)P_{LO}}\cos{\Delta \varphi}=2R(\lambda)GCE_s(t)E_{LO}\cos{\Delta \varphi},
\end{align}
where $R(\lambda)$ is the responsivity of photodiodes;  $G$ is the
electronic gain of balanced detector; \textcolor{black}{$C=S/(2\xi)$
  is the ratio of the effective beam area $S$ and the
  doubled impedance $\xi$ of the medium;
  $E_s$ and $E_{LO}$ are the amplitudes of the initial signal field and
  the LO, respectively}.

In the case of heterodyning with $\omega_{s}\ne \omega_{LO}$,
the intensities of
the mix of two harmonic signals 
contain harmonics oscillating
at the difference
and
the sum frequencies:
$\omega_-=\omega_{s}-\omega_{LO}$ and
$\omega_+=\omega_{s}+\omega_{LO}$.
Since both
the optical and the sum frequencies are much larger than
the bandwidth of any existing photodiode,
the photodetectors are unable to register these high frequency harmonics.
Therefore, the output of the balanced
detector will be the electrical signal of the difference frequency $\omega_-$
that stores the information about the phase of the initial
signal~\cite{yoshida200864,hongo20071}.
% regardless of guessing it from the receiver, and it can be obtained
% from the analysis of the final harmonic
% function
This approach allows
detecting two  quadrature components at the once, since the receiver
now does not need to set any phase of the local oscillator.
The output photocurrent is given by 
\begin{align}
\label{currentref2}
     I(t) =2R(\lambda)G\sqrt{P_s(t)P_{LO}}\cos{(\omega_-t+\Delta \varphi)}.
\end{align}

%%%%%%%%%%%%%%%%%%%%%%%%%%%%%%%%%%%%
\subsection{Coherent detection with single quadrature selection}
\label{subsec:scw-1}
%%%%%%%%%%%%%%%%%%%%%%%%%%%%%%%%%%%%%%

\begin{figure}[ht!]
\centering
\includegraphics[width=\textwidth]{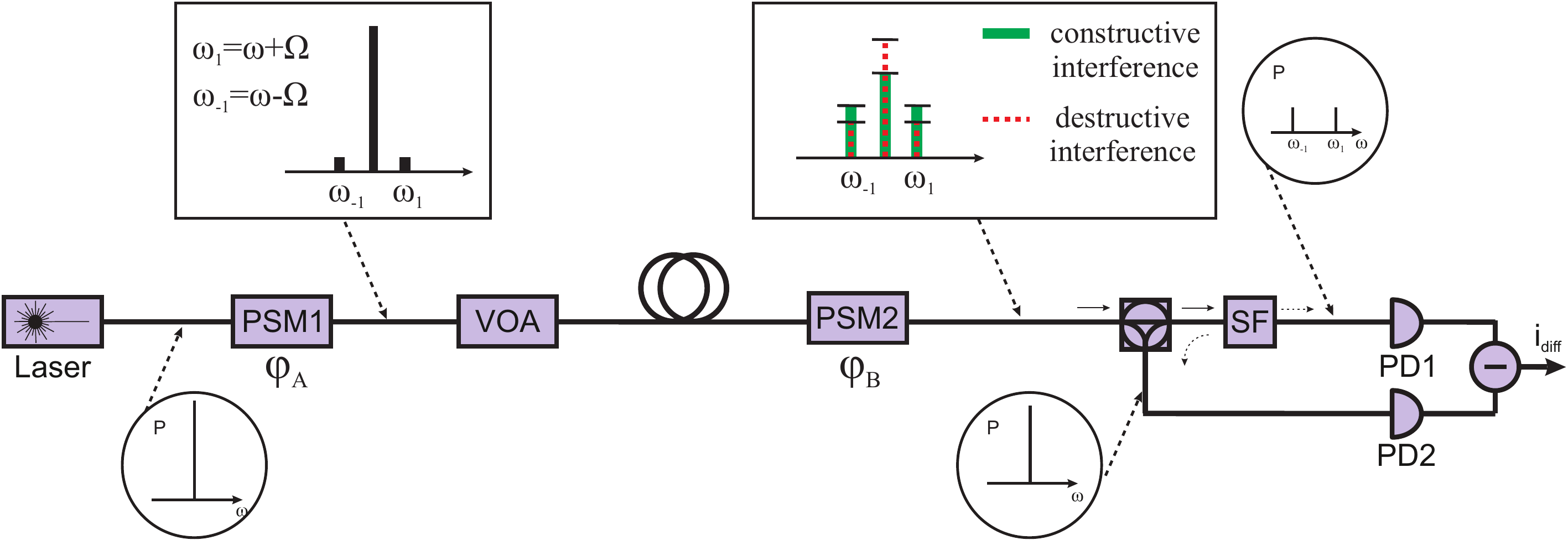}
\caption{Principal scheme
  of SCW CV-QKD setup. PSM is the electro-optic phase modulator; VOA
  is the variable optical attenuator; SF is the spectral filter that cuts
  off the carrier; PD is the photodiode. Diagrams in circles show the
  simplified power spectrum, diagrams in squares illustrate the
  spectrum for various phase shifts.
  Horizontal dashes mark the levels of intensities and are added for
  illustrative purposes.}  
\label{fig2}
\end{figure}

In this section
we describe the coherent detection scheme with single quadrature
selection for the  SCW QKD system.
We keep the Alice's block identical
to the original sender block
of the system~\cite{gleim2016secure}
and change the block of Bob
by replacing the single photon detector
with the balanced detector.
Figure~\ref{fig2} details the changes and the mode of operation of
the proposed scheme. 

% comment
% By definition, homodyne detection is characterized by interference
% of a weak signal with a powerful local oscillator on a $50/50$ beam
% splitter. After interference, the mean number of photons at the
% photodiodes $n_1$ and $n_2$ depends on parties' phase difference
% $\Delta\varphi=\varphi_{a}-\varphi_{b}$. Then, the difference in
% photo-electrons $n_e$ can be determined by subtracting the received
% signal values.  

Coherent detection illustrated in Fig.~\ref{fig2} 
is functionally similar to conventional
homodyne detection described in the previous section.
By contrast to the homodyne detection scheme,
the phase modulator in the Bob's module
plays the role of the $50/50$ beam splitter.
After the second modulation the interference is
observed at frequencies $\omega_{k}=\omega+k\Omega$
provided that Alice and Bob use
the microwave modulating field with identical frequencies.

In other words,
the local oscillator is not used directly as a separate
source.
SCW homodyning occurs as a result of
redistribution of the energy
between the intense carrier mode and the weak sidebands,
as if a strong coherent
beam is mixed with them at a beam splitter. 

\begin{figure}[ht!]
\centering
\includegraphics[width=\textwidth]{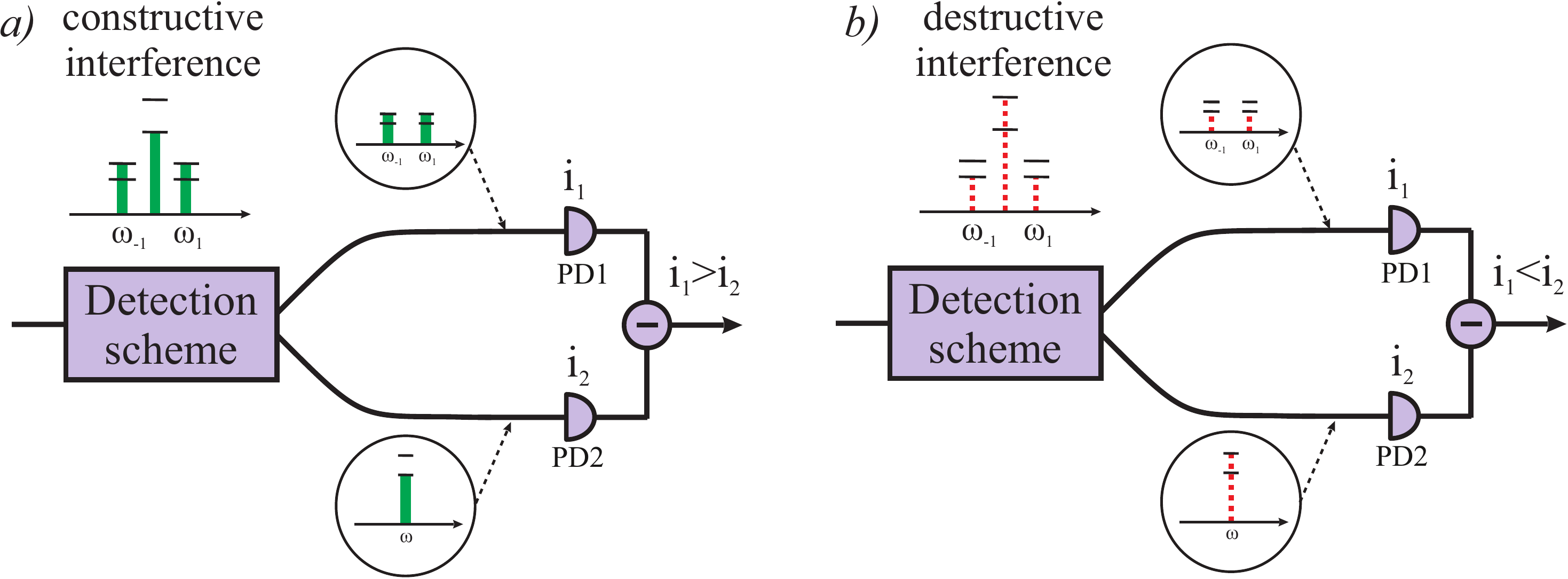} 
\caption{SCW coherent detection scheme. The charts show
  the energy distribution between the carrier and the subcarriers
  for (a) constructive and (b) destructive interference.
  % The power of subcarrier
  % signal is higher or lower than the carrier power,
  % respectively.
} 
\label{fig3}
\end{figure}

As is illustrated in
Fig.~\ref{fig3}a (\ref{fig3}b),
the power of subcarrier wave is
higher (lower) than the carrier wave power
when the interference is constructive (destructive).
A narrow spectral filter then separates the carrier from the
sidebands. Finally, the outputs (the carrier and the sidebands) are
detected by two different photodiodes, and their photocurrents are
subtracted. Thus, one can extract information encoded in the
oscillating signal phase.
Similar to the conventional homodyne detection
in QKD, Bob measures only one quadrature component at a time by
selecting the phase of the microwave field from the set of
$\varphi_b\in\{0,\:\pi/2\}$. 

% \subsection{Detection model}
\label{sec:1.1}

Now we describe
a simple classical model
(quantum considerations
can be found in~\cite{samsonov2019continuous})
of the above homodyne-like coherent detection scheme.
% u the carrier wave and the subcarriers
% as the reference and signal waves, repectively.
The traveling wave phase modulator on the Bob's side
generally differs from  the Alice's modulator
both in the modulation index,
$m_b\ne m_a$,
and in
the phase, $\varphi_b\ne\varphi_a$, 
thus introducing the phase difference $\Delta \varphi=\varphi_a-\varphi_b$.
The expression for the
output field emerging after the second modulator $E'(t)$
\begin{equation}
  \label{eq:E-mod-1}
    E'(t) = E_0e^{i\omega t}e^{i m_a \cos(\Omega t+\varphi_{a})}e^{i m_b \cos(\Omega t+\varphi_{b})}
\end{equation}
can be simplified with the help of identities
\begin{subequations}
\begin{align}
  &
  \label{eq:m1}
m_{a} \cos \left(\Omega t+\varphi_{a}\right)+m_{b} \cos \left(\Omega
    t+\varphi_{b}\right)=m \cos (\Omega t+\phi),
  \\
  &
    \label{eq:m2}
m=\sqrt{m_{a}^{2}+m_{b}^{2}+2 m_{a} m_{b} \cos
    \left(\varphi_{a}-\varphi_{b}\right)},
  \\
  &
    \label{eq:m3}
    m \mathrm{e}^{i\phi}=m_a\mathrm{e}^{i\varphi_a}+m_b \mathrm{e}^{i\varphi_b}.    
\end{align}
\end{subequations}
%
% \begin{equation}
% \begin{aligned}
% E'(t) = E_0e^{i\omega t}e^{i m_a \cos(\Omega t+\varphi_{a})}e^{i m_b \cos(\Omega t+\varphi_{b})}\\
% =E_0e^{i\omega t}\exp\left(im_a\left[\cos{\left(\Omega t +\frac{\varphi_a+\varphi_b}{2}\right)}\cos{\frac{\Delta \varphi}{2}}-\sin{\left(\Omega t+ \frac{\varphi_a+\varphi_b}{2}\right)}\sin{\frac{\Delta \varphi}{2}}\right]\right.\\
% \left.+im_b\left[\cos{\left(\Omega t +\frac{\varphi_a+\varphi_b}{2}\right)}\cos{\frac{\Delta \varphi}{2}}+\sin{\left(\Omega t+ \frac{\varphi_a+\varphi_b}{2}\right)}\sin{\frac{\Delta \varphi}{2}}\right]\right)\\
% =E_0e^{i\omega t}\exp\left[i(m_a+m_b)\cos{\frac{\Delta \varphi}{2}}\cos{\left(\Omega t +\frac{\varphi_a+\varphi_b}{2}\right)}\right.\\
% \left.+i(m_b-m_a)\sin{\frac{\Delta \varphi}{2}}\sin{\left(\Omega t +\frac{\varphi_a+\varphi_b}{2}\right)}\right],
% \end{aligned}
% \end{equation}
%
Similar to Eq.~\eqref{e1},
the output field can now be written in
the form of the Jacobi-Anger expansion: 
\begin{equation}
  \label{eq:E-mod}
E'(t)=E_0e^{i\omega t} \sum_{k=-\infty}^{\infty} i^{k} J_{k}(m)
\mathrm{e}^{i k \phi} \mathrm{e}^{i k \Omega t} \equiv
\sum_{k=-\infty}^{\infty} E_{k} \mathrm{e}^{i k \Omega t}, \quad
E_{k}=E_0e^{i\omega t} i^{|k|} J_{|k|}(m) \mathrm{e}^{i k \phi}. 
\end{equation}
The carrier wave is separated from the sidebands
using spectral filtering in the Bob’s module
For simplicity,
we shall neglect the losses and imperfection of the
spectral filter.
So, we can single out
the component at the central frequency
to obtain the following expressions for
the fields in two arms of the detector:
\begin{align}
  &
    \label{eq:E-12}
  E_1(t)  = E_0e^{i\omega t} J_{0}(m),
  \quad
E_2(t) = \sum_{\substack{k=-\infty,\\k \neq 0}}^{\infty} E_{k} \mathrm{e}^{i k \Omega t}.
\end{align}
Now we can
closely follow the line of reasoning
led to Eq.~\eqref{currentref},
and obtain
the time averaged difference of the photocurrents
in the form: 
\begin{align}
  &
    \label{eq:I-hom-1}
    I = R(\lambda)GC\langle |E_2(t)|^2 - |E_1(t)|^2\rangle_t
    =R(\lambda)G C E_0^2(1 - 2 J_0^2(m)),
\end{align}
where
$\displaystyle\langle\ldots\rangle_t=\tau^{-1}\int_0^\tau\ldots\mathrm{d}
t$,
$\tau=2\pi/\Omega$.
Using temporal averaging assumes that
the detectors are insensitive to
harmonics of intensities
$|E_1(t)|^2$ and $|E_2(t)|^2$
oscillating at radio frequencies.
Note that modulation keeps the amplitude of the phase-modulated
wave~\eqref{eq:E-mod-1}
unchanged with $|E'(t)|^2=E_0^2$,
so that
Eq.~\eqref{eq:E-mod}
leads to
the unitarity condition
$\displaystyle\sum_{k=-\infty}^{\infty} J_{k}^2(m)=1$
used in derivation of
the right-hand side of formula~\eqref{eq:I-hom-1}.

When the phase difference
$\Delta\varphi=\varphi_{a}-\varphi_{b}$
is changed,
the modulation index $m$ varies between
its minimal value
$m_{\mathrm{min}}=|m_b-m_a|$
at $\cos(\varphi_{a}-\varphi_{b})=-1$
to the maximal index of modulation
$m_{\mathrm{max}}=m_b+m_a$
at $\cos(\varphi_{a}-\varphi_{b})=1$.
In the case where Alice sends non-modulated wave
with $m_a=0$,
the output voltage~\eqref{eq:I-hom-1}
will be zero provided
the Bob's index of modulation is adjusted
to meet
the condition:
$J_0(m_b)=1/\sqrt{2}$.

% In this way, the observed process of interference, marked by the flow
% of energy from the strong carrier mode to the sidebands, fully
% corresponds to the usual mixing of two signals oscillating at one
% frequency on a beam splitter. 

\begin{figure}[ht]
\centering
\includegraphics[width=0.6\linewidth]{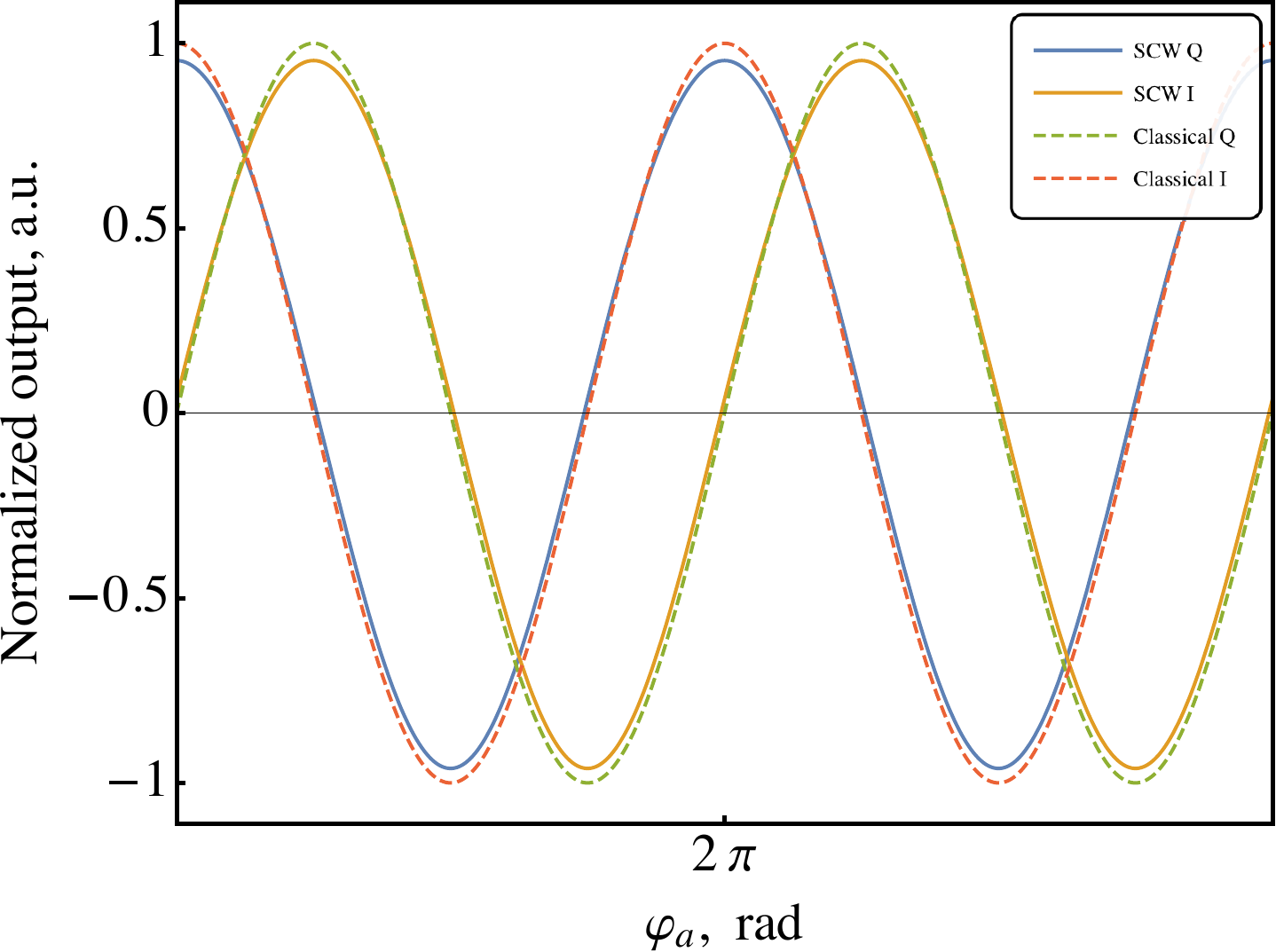} 
\caption{
  The in-phase and the quadrature components of
  the normalized output
  signals from the SCW coherent detection scheme
(see Eq.~\eqref{eq:I-norm})
  and the classical homodyne
  scheme.
} 
\label{fig4}
\end{figure}

Figure~\ref{fig4} shows
the curves representing
dependence of the output voltage
on the phase shift $\varphi_a$
for the in-phase component $I$ and the
the quadrature component $Q$
that are computed at $\varphi_b = 0$ and
$\varphi_b = \pi/2$, respectively.
In our calculations,
the parameters
are taken to be close
to those used in the experiments
verifying our model.
These are:
the
carrier wave power is
$P_{c}=10$ $\mu$W;
the modulation index of Alice is
$m_a = 0.09$;
the responsivity of the photodiodes
is $R= 0.6$; 
and
the additional gain of the detectors
is $G=4\times 10^3$.s 

In order to draw analogy
between the proposed
scheme and
conventional homodyne detection,
we assume that
the temporally averaged
intensities
$\langle |E_1(t)|^2\rangle_t$
and
$\langle |E_2(t)|^2\rangle_t$
of the carrier and
the subcarrier
waves at the output of Alice's modulator
correspond to
the intensities of the signal
and reference waves:  
$E_s=E_0\sqrt{1-J_0^2(m_a)}$
and
$E_{LO}=E_0 J_0(m_a)$.
Then, for the classical homodyne scheme,
the curves for the normalized output voltage~\eqref{currentref}
$I/I_{\mathrm{max}}=\cos(\Delta\varphi)$
are plotted in Figure~\ref{fig4}.
It can be seen that the normalized output signal~\eqref{eq:I-hom-1}
for our scheme
\begin{align}
  \label{eq:I-norm}
  I/I_{\mathrm{max}}=\frac{1 - 2 J_0^2(m)}{2 J_0(m_a)\sqrt{1-J_0^2(m_a)}}
\end{align}
agrees closely
with the predicitions of the classical homodyne scheme. 

%%%%%%%%%%%%%%%%%%%%%%%%%%%%%%%%%%%%%%%
\subsection{Simultaneous selection of both quadrature component}
\label{subsec:scw-2}
%%%%%%%%%%%%%%%%%%%%%%%%%%%%%%%%%%%%%%%

\begin{figure}[ht]
\centering
\includegraphics[width=\textwidth]{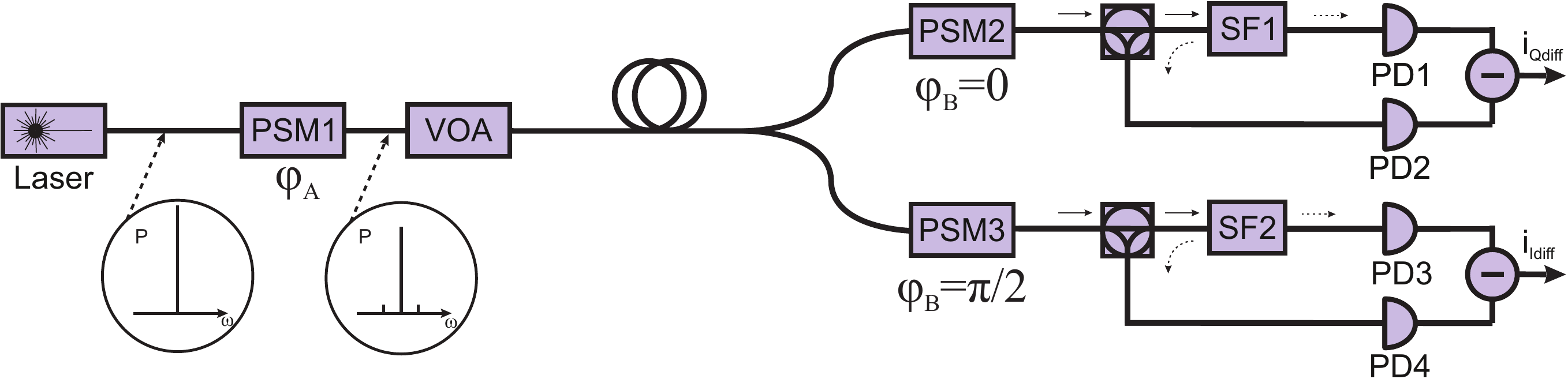} 
\caption{Principal scheme of SCW CV-QKD setup with phase-diversity coherent receiver.}
\label{fig6}
\end{figure}

An important point is that
it is possible to construct
an optical phase-diversity coherent detection scheme.
In this scheme,
the receiver is similar to $90^\circ$ optical hybrid and
both the quadrature components can be measured simultaneously.
The principal
scheme of SCW CV-QKD setup utilizing such a receiver is shown in
Figure~\ref{fig6}.
In this setup,
the Y beam splitter is combined with the two coherent
detection schemes where the electro-optic phase modulators
are displaced in phase by $90^\circ$.  

By using the detection scheme depicted in Figure~\ref{fig6},
we can
measure
the two output signals, $E_{1I}(t)$ and $E_{2I}(t)$, from the top arm of the
detector with the phase $\varphi_b = 0$,
whereas the output signals, $E_{1Q}(t)$ and $E_{2Q}(t)$,
from the bottom arm emerge from the detector with the phase $\varphi_b = \pi/2$.  
Then the output currents after the balanced detectors are
converted to the voltages $V_{I,Q}$.
Hence by using these two balanced detectors
we can perform simultaneous I/Q measurements.  
Note, however, that this approach will
introduce an additional 3~dB loss at the detection
stage and its QKD security should be carefully analyzed. 

%%%%%%%%%%%%%%%%%%%%%%%
\subsection{Heterodyne detection}
\label{subsec:heterodyne}
%%%%%%%%%%%%%%%%%%%%%%

\begin{figure}[ht!]
\centering
\includegraphics[width=\textwidth]{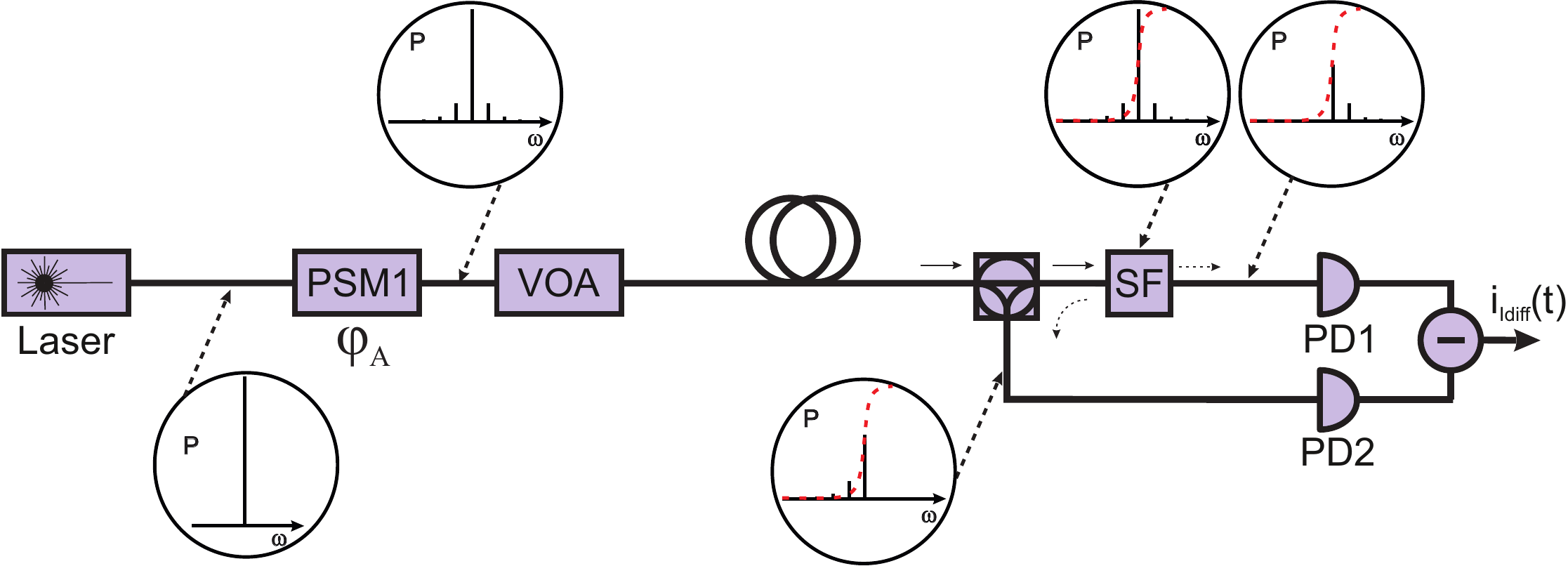} \caption{Principal scheme
  of SCW CV-QKD setup with heterodyne detection.
Examples of power spectra are shown in encircled diagrams
  and
$i_{\mathrm{diff}}$ stands for the output  photocurrent.}  
\label{fig8}
\end{figure}

Another well-known method
to obtain information about
the optical complex amplitude
is heterodyne
detection~\cite{protopopov,kikuchi2015fundamentals,gebbie1967heterodyne,maznev1998optical,fried1967optical,delange1968optical}.
In this section 
we show how a heterodyne-type detection scheme
can be adopted for quadrature phase-coded multimode signals.
The mode of operation of our coherent detection
scheme
demonstrating versatility of the SCW method
is described in Figure~\ref{fig8}.
Note also that such scheme promises
significant simplification of implementation.

As is shown in Figure~\ref{fig8},
similar to the scheme presented in Figure~\ref{fig2},
the only component changed in
the initial system~\cite{gleim2016secure,miroshnichenko2018security}
is
the Bob's block whose
detection scheme is no longer using the phase modulator.
In this scheme
spectral filtering is used to mix
the carrier wave with
the upper and the lower sidebands
where
subcarrier frequencies are higher and below
the carrier frequency $\omega$.
% to down-convert the signal, including
% the amplitude and the phase information, to the intermediate
% frequency
For this purpose, the spectral filter bandpass is
chosen so as
to pass light at the frequency $\omega + \Omega$
and to halve the intensity of the carrier
  at the frequency $\omega$,
whereas the rest of light is reflected back.
  In other words, the transparency of the spectral filter, $T$,
  should meet the conditions:
  $T(\omega) \approx 0.5$, $T(\omega-\Omega) \to 0$  and
  $T(\omega+\Omega) \to 1$.

  From Eq.~\eqref{e1}
  we obtain
  the fields in the separate arms
  in  the following form:
\begin{equation}
\label{e7}
E_1(t)  = E_0e^{i\omega t}(\sum_{k=1}^{\infty}{i^kJ_k(m_a)e^{ik(\Omega t+\varphi_{a})}}+\sqrt{0.5}J_0(m_a)),
\end{equation}
\begin{equation}
\label{e8}
E_2(t)  = E_0e^{i\omega t}(\sum_{k=1}^{\infty}{i^kJ_k(m_a)e^{-ik(\Omega t+\varphi_{a})}}+\sqrt{0.5}J_0(m_a)),
\end{equation}
where we have used the indentity
$J_{-\alpha}(x)=(-1)^{\alpha} J_{\alpha}(x)$ for the Bessel functions.
When the modulation index $m_a$ is small,
we can apply the lowest order approximation
that takes into account
only the first order subcarriers
\begin{equation}
E_1(t)  \approx E_0(iJ_1(m_a)e^{i(\omega_{+}t+\varphi_a)}+\sqrt{0.5}J_0(m_a)e^{i\omega t}),
\end{equation}
\begin{equation}
E_2(t)  \approx E_0(iJ_1(m_a)e^{i(\omega_{-}t-\varphi_a)}+\sqrt{0.5}J_0(m_a)e^{i\omega t}),
\end{equation}
where $\omega_{\pm}=\omega \pm \Omega$,
and derive the output photocurrent in the absence of
  noise 
\begin{equation}
\label{e11}
    I(t) = R(\lambda)GC (|E_{2}(t) |^2-|E_{1}(t) |^2) = R(\lambda)GC 2 \sqrt{2} E_{0}^2 J_0(m_a) J_1(m_a) \sin(\Omega t+\varphi_a).
\end{equation}
Clearly, this result bears a striking resemblance
to the classical heterodyning.  

\begin{figure}[ht]
\centering
\includegraphics[width=0.6\linewidth]{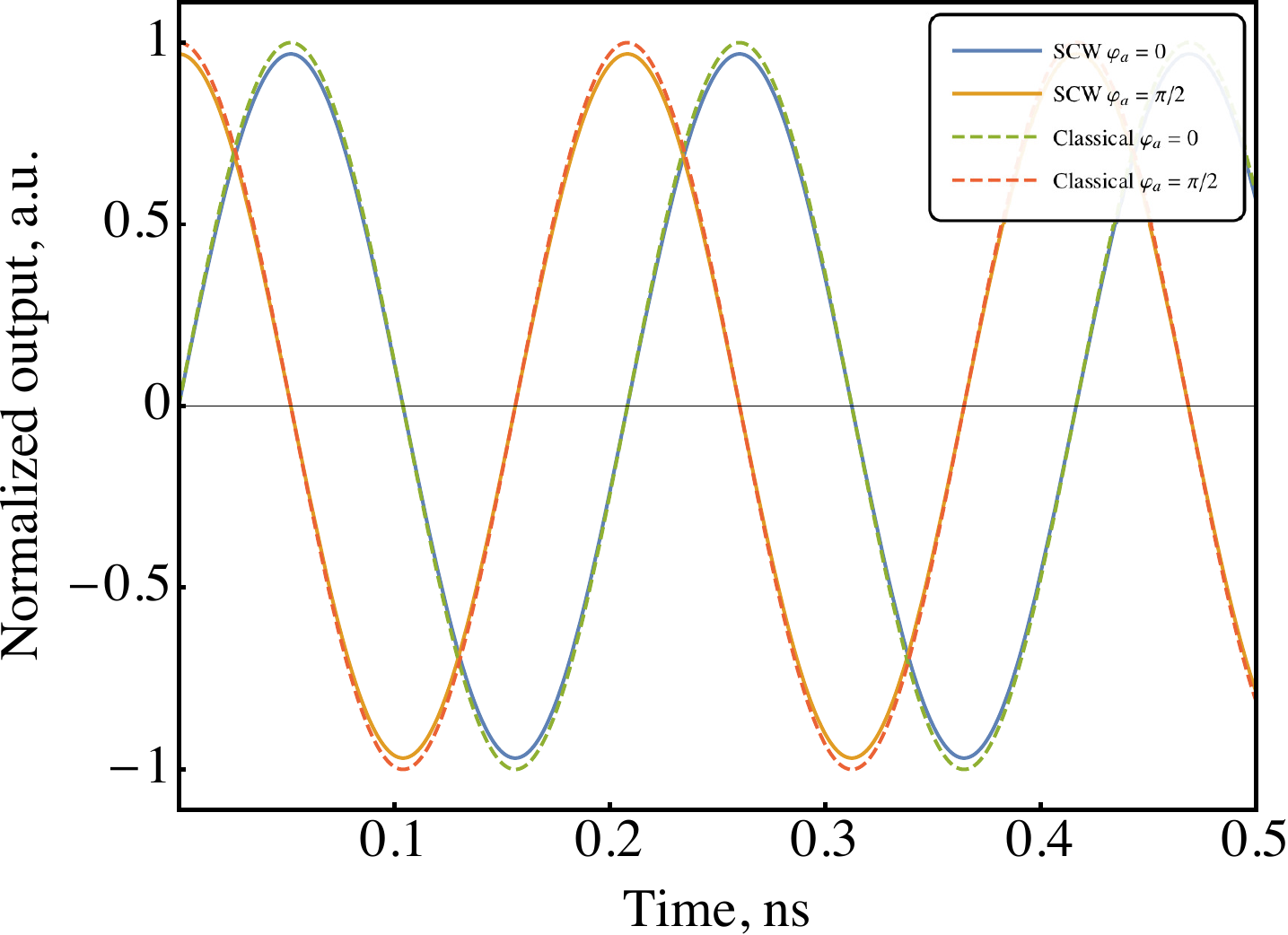} 
\caption{
  Time dependence of the normalized heterodyne-derived signals
  for the SCW and the classical detection schemes.
} 
\label{sine}
\end{figure}

As in the previous section,
we now discuss this analogy
using
Eq.~\eqref{currentref2}
with $\omega$ and $\Delta\varphi$
replaced by
$\Omega$ and $\varphi_a+\pi/2$,
respectively.
For this purpose,
we assume that
the amplitudes of the signal
and reference waves are  
$E_s=E_0\sqrt{1-J_0^2(m_a)}$
and
$E_{LO}=E_0 J_0(m_a)$
and compare
the output
of the classical heterodyne,
$I_{\mathrm{max}}\sin(\Omega t+\varphi_a)$,
with the signal given in Eq.~\eqref{e11}.
Figure~\ref{sine} shows
that the normalized output
of our scheme
\begin{align}
  \label{eq:e11-norm}
  I(t)/I_{\mathrm{max}}=\frac{\sqrt{2}J_1(m_a)}{\sqrt{1-J_0^2(m_a)}}\sin(\Omega t+\varphi_a)
\end{align}
turns out to be very close to the
output of the classical scheme.
It should be stressed that, in this analysis,
the detectors are assumed to be sensitive to the
radio frequency $\Omega$.

% igure~\ref{sine} shows the dependence of output on the time for the
% proposed scheme and for the classical heterodyne in which the LO with
% the carrier wave amplitude at the frequency $\omega$ mix with the signal
% with amplitude of all subcarriers at the frequency $\Omega$ on a 50/50
% beam splitter.
% Here we assume that the detectors are
% For calculations we use the same parameters
% as in the section~\ref{sec:1}.

Demodulation of the received signal can be carried out by standard
telecommunication methods using a quadrature synchronous
demodulator~\cite{kikuchi2015fundamentals,bylina2017comparative,wang2015iq,bohme1999heterodyne,chen2010optical}.
The demodulator setup allows one to observe the in-phase (initial) and
quadrature (shifted by $\pi/2$ using a voltage generator) signal
components simultaneously on the analyzer and thus extract the information
on the phase selected by Alice.  

% The main disadvantage of the proposed heterodyne detector for SCW-QKD system is the necessity to deal with a high intermediate frequency ($\Omega=4.8$~GHz), contraction of frequency $\Omega$ leads to difficulty in spectral filtering. 
The main disadvantage of this method is emerging conflict between
spectral filtering and bandwidth of balanced detector.
On the one hand, an
increase of modulation frequency $\Omega$ leads to the growth of
required detector's bandwidth, what results in extra electronic
noises~\cite{chi2011balanced}. Usually bandwidth of balanced detectors
used for quantum measurements is less than
1~GHz~\cite{tang2020performance}.
On the other hand, low modulation
frequencies perplex subcarriers' spectral filtering.
Thus one of the
main challenges for practical implementation of such scheme is to find
an optimal modulation frequency that satisfies both of these
conditions. 

% \textcolor{black}{For all schemes, it should also be mentioned that
%   the use of too large modulation index, and, as a consequence, too
%   high power of the initial signal, will lead to the fact that a
%   significant amount of the energy will be transferred to the
%   sidebands of higher orders, and the output voltage will be much less
%   even without taking into account losses.} 

%%%%%%%%%%%%%%%%%%%%%%%%
\section{Experimental results}
\label{sec:experimental}
%%%%%%%%%%%%%%%%%%%%%%%%

In this section we present
the results of our experiments
verifying the theoretical models
presented in the previous section.
For the homodyne-like detection setup shown in
Figure~\ref{fig2},
our proof-of-principle experiment
have used
a 1550~nm 10~$\mu$W fiber-coupled laser
directed into the LiNbO$_3$
electro-optic phase modulator with the electrical signal frequency
$\Omega = 4.8$~GHz.
The modulation index $m_b$ is adjusted to maximize the difference
between the results for different values of $\Delta\varphi$
and the phase dependence of the output voltage is expected
to be nearly harmonic (see Figure~\ref{fig4}).
After
the second modulation the spectral components are transmitted through
the circulator to a fiber Bragg grating spectral filter.
The two
output ports of the circulator are coupled to the input ports of a
self-developed balanced detector
(the measurement bandwidth is 100~MHz, the gain is
$G = 4 \cdot 10^3$ and the responsivity is $R = 0.6$). 

\begin{figure}[ht]
\centering
\begin{subfigure}{.5\textwidth}
  \centering
  \includegraphics[trim=70 0 60 0,width=\linewidth]{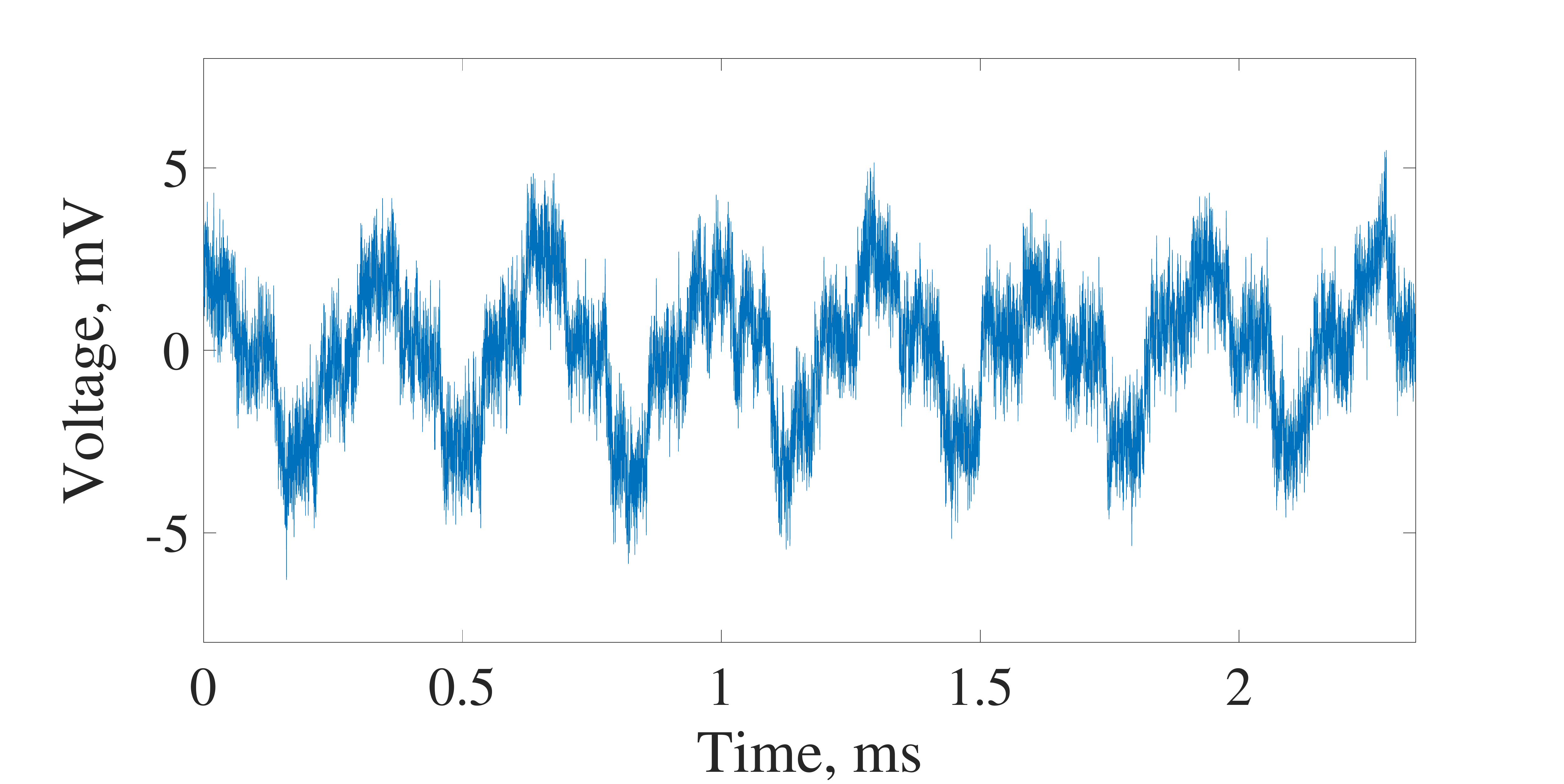}
  \caption{}
  \label{fig50}
\end{subfigure}%
\begin{subfigure}{.5\textwidth}
  \centering
  \includegraphics[trim=70 0 60 0,width=\linewidth]{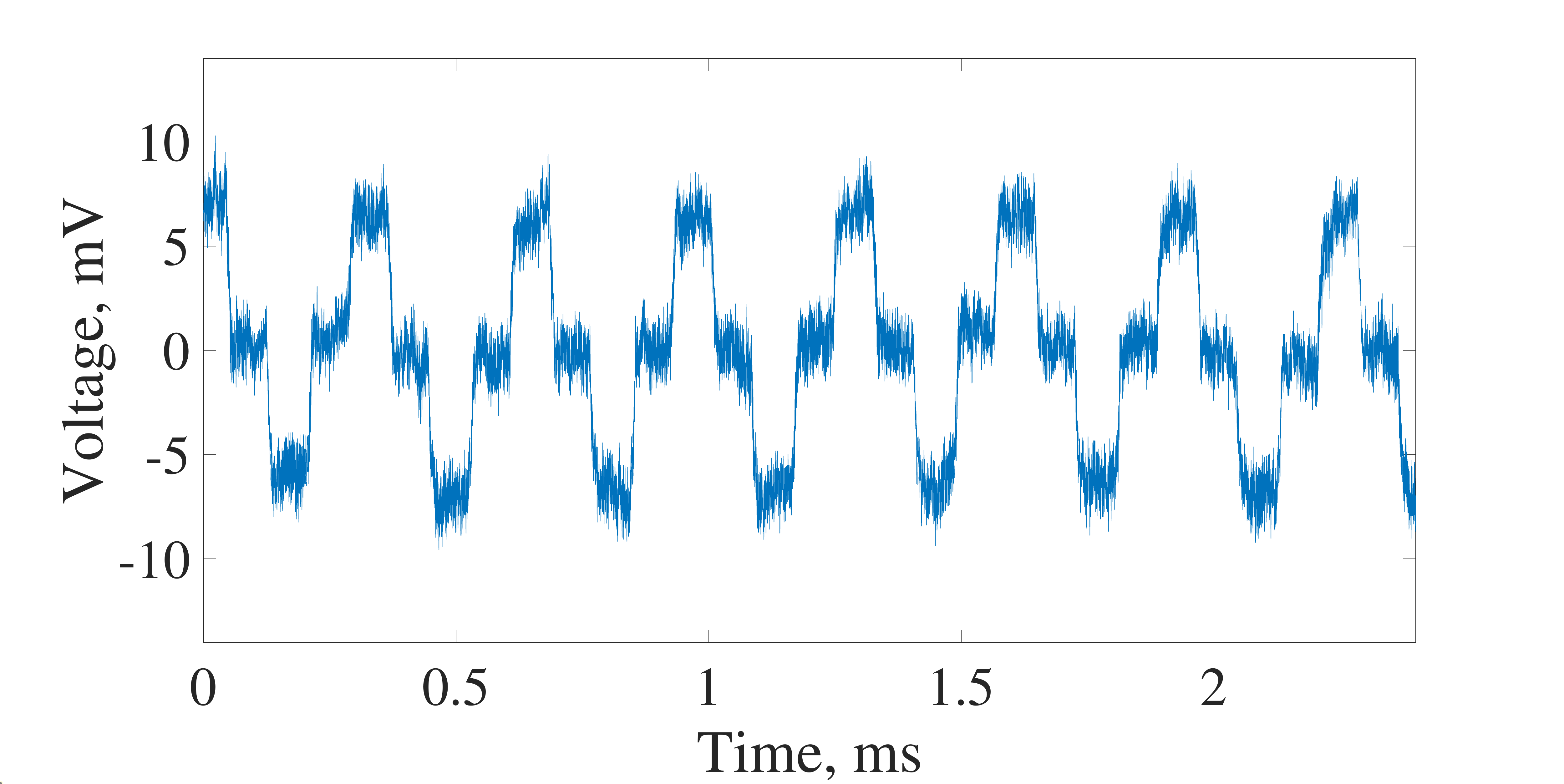}
  \caption{}
  \label{fig51}
\end{subfigure}
\caption{Time dependence of the output voltage for
  the homodyne-like coherent detection scheme shown in Fig.~\ref{fig2}.
  The extreme points correspond to the
  constructive and destructive interference. The phase difference
  $\Delta \varphi$ varies periodically in time
  switching between the values from a discrete set:
  $\{0,\pi/2,\pi,3\pi/2\}$.
  Two cases are shown: (a)~$P_c =
  9.96$~$\mu$W, $P_s = 40.00$~nW; and (b)~$P_c = 9.5$~$\mu$W, $P_s
  = 500.00$~nW.} 
\label{fig5}
\end{figure}

Figure~\ref{fig5} presents the
output voltage measured in relation
of time when the phase difference
(the phase $\varphi_a$ is set to be zero),
$\Delta\varphi$,
undergoes time-periodic
piecewise changes between
the states with
equidistant values from
a discrete set $\{0,\pi/2,\pi,3\pi/2\}$.
Referring to Fig.~\ref{fig50},
it can be seen that,
in  the case where
the carrier wave power after the modulation
was $P_c = 9.96$~$\mu$W and
the total power of the subcarriers was $P_s = 40.00$~nW,
variations of the voltage level are almost indiscernible.
By contrast,
Figure~\ref{fig51} demonstrates
the case with
unambiguously distinguishable voltage levels
that occurs at
$P_c = 9.5$~$\mu$W and $P_s = 500.00$~nW.
Thus,
our coherent detection scheme
is sufficiently sensitive to estimate the phase of the
quadrature phase-coded multimode signals from the electro-optic
phase modulator.

\begin{figure}[ht]
\centering
\includegraphics[width=0.45\textwidth]{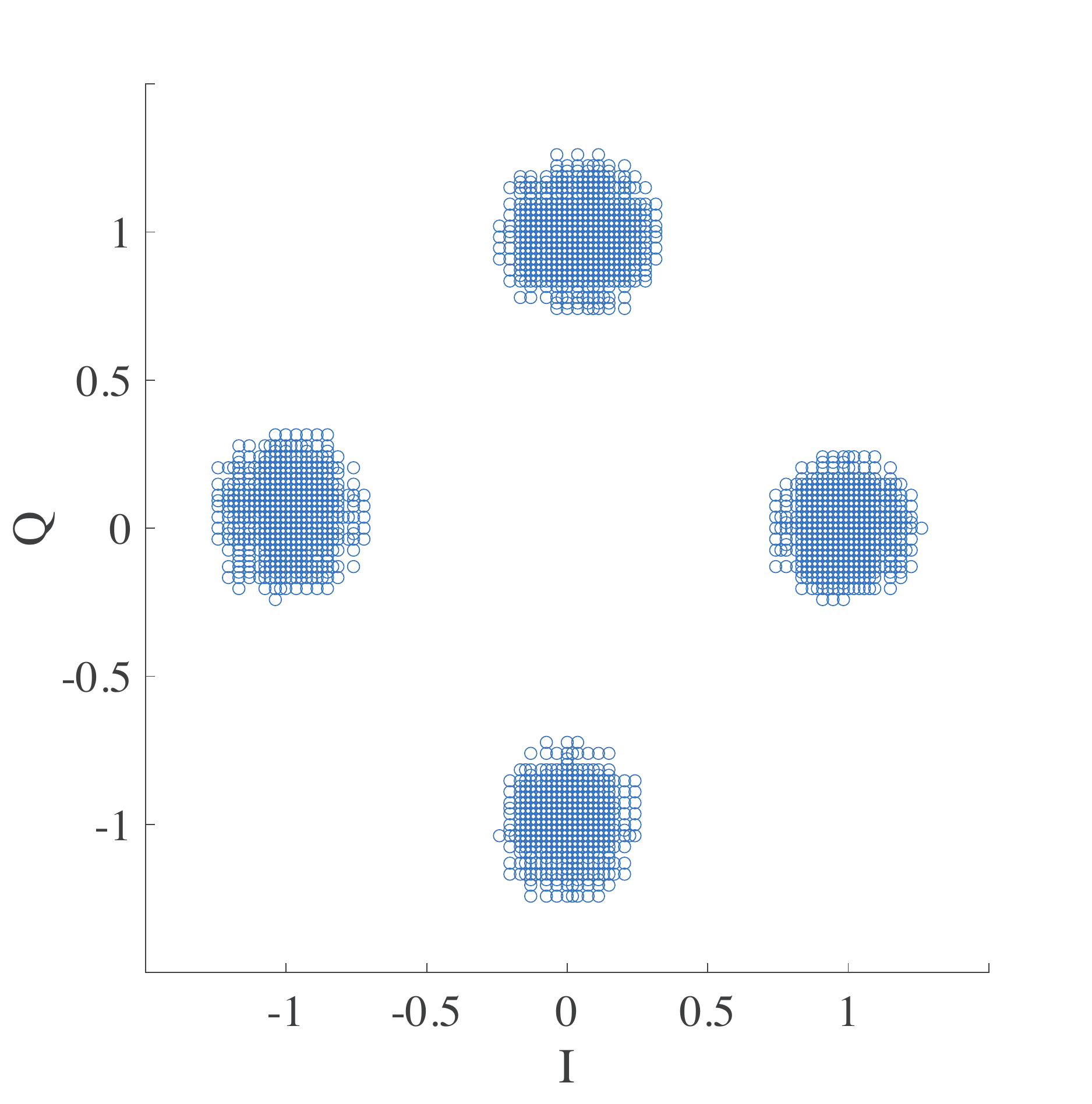} 
\caption{4-PSK constellation diagram recovered from the
  phase-modulated optical signal, $P_c = 9.5$ $\mu$W, $P_s = 500.00$
  nW.} 
\label{fig7}
\end{figure}

We have also performed the experiment
for the optical phase-diversity coherent receiver
sketched in Figure~\ref{fig6}.
The 4-PSK constellation diagram
recovered from the phase-modulated optical signal using the optical
phase-diversity coherent receiver is depicted in Figure~\ref{fig7}. 

\begin{figure}[h!!!]
\centering
\includegraphics[width=0.6\textwidth]{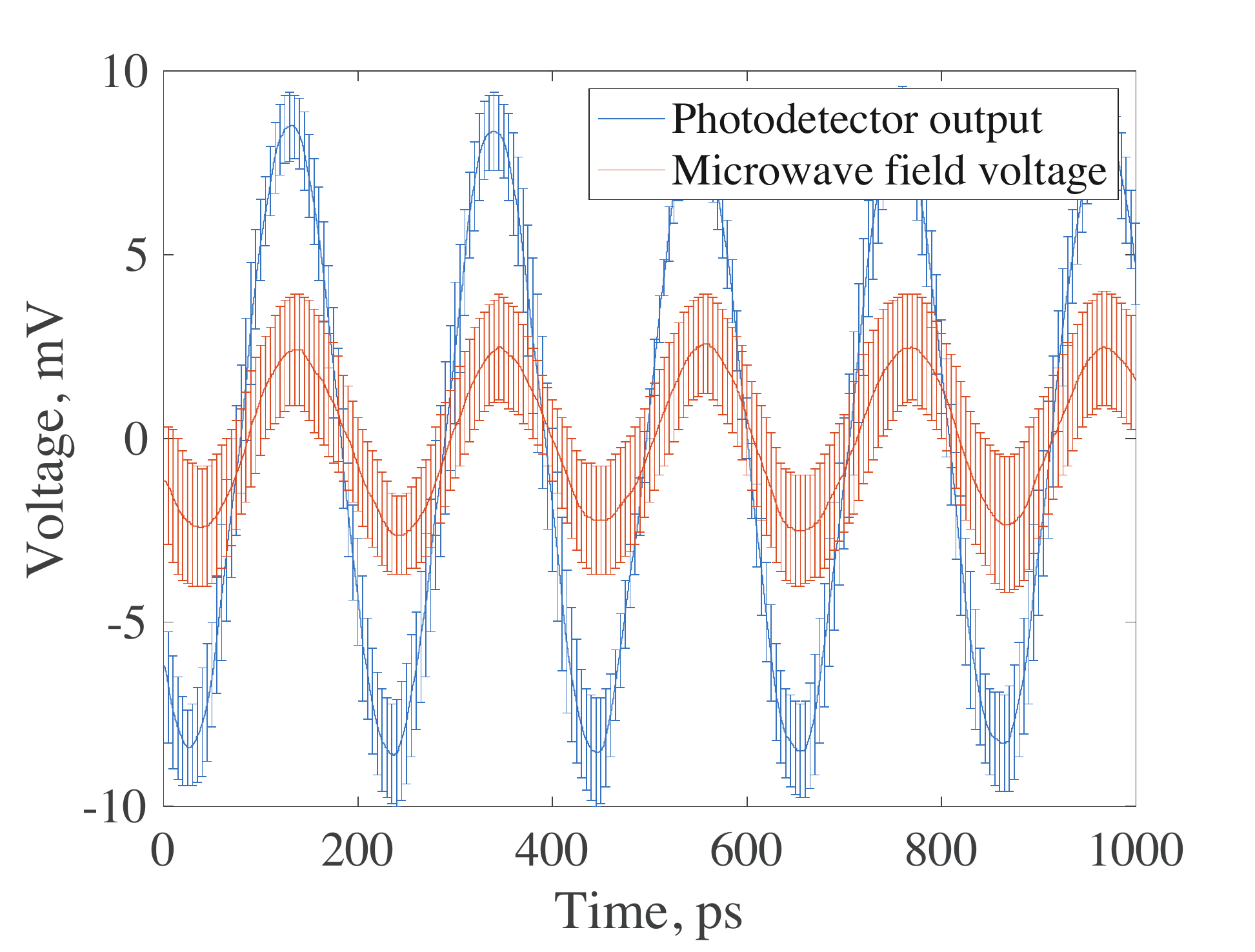} 
\caption{Time dependence of microwave field voltage and the output
  voltage for balanced detector's
  single arm ($P_c = 225.00$~$\mu$W, $P_s = 146.50$~$\mu$W) using the
  SCW heterodyning.
  Mean values are indicated along with standard deviations.}  
\label{fig10}
\end{figure}
In order to test the heterodyne scheme (see Figure~\ref{fig8}),
we  need to get around the
above-mentioned difficulties concerning the frequency bandwidth.
For this purpose,
we have utilized the photodetector with 6.17~GHz
bandwidth in the single arm of the scheme described in Figure~\ref{fig8}
to obtain the signal at the intermediate frequency.
The measured curves
for the output voltage and the microwave field
are presented in Figure~\ref{fig10}.
The results show that
the signal light is down-converted to
the intermediate frequency coincident with
the frequency of
the microwave field used in the
electro-optic phase modulator. 

%%%%%%%%%%%%%%%%%%%%%%
\section{Conclusion}
\label{sec:conclusion}
%%%%%%%%%%%%%%%%%%%%%%

In this paper we
have applied several approaches
to put coherent detection schemes
into the framework of
the subcarrier wave QKD systems.
In our homodyne-like coherent detection schemes
(see Sec.~\ref{subsec:scw-1} and Sec.~\ref{subsec:scw-2}),
the phase-modulated light
emerging from the Bob's modulator   
is splitted into
the carrier and the subcarrier waves
and
the output signal is determined by
the difference between
the responses of the photodiodes
to the light of these two waves.
By contrast,
in our heterodyne-type detection scheme
described in Sec.~\ref{subsec:heterodyne}
the light modulated by Alice
is directly used as the input wave.
This wave, similar to the homodyne-like scheme,
is divided into two waves. But,
in contrast to this scheme,
these waves are obtained by
filtering out either the lower or
the upper sidebands.
In this case, we have shown that
the differential signal which is oscillating with
the microwave frequency
can be used to extract
the information about the phase encoded by Alice.

In our theoretical considerations
we have presented simple classical models
emphasizing the analogy
between the suggested coherent detection schemes
and the standard
homodyne/heterodyne methods.
Our experimental results
clearly demonstrate
practical feasibility of the detection schemes. 
Futher development will require
a more sophisticated and detailed 
analysis of these schemes
using the methods of quantum optics.
This will allow their legitimate employment in quantum
computing, quantum cryptography, and quantum tomography.

\ack
This work was funded by Government of Russian Federation (Grant No. MK-777.2020.8).

%%%%%%%%%%%%%%%%%%%%%%% References %%%%%%%%%%%%%%%%%%%%%%%%%
\section*{References}
\bibliographystyle{iopart-num}
\bibliography{main.bib}

\providecommand{\newblock}{}
\begin{thebibliography}{10}
\expandafter\ifx\csname url\endcsname\relax
  \def\url#1{{\tt #1}}\fi
\expandafter\ifx\csname urlprefix\endcsname\relax\def\urlprefix{URL }\fi
\providecommand{\eprint}[2][]{\url{#2}}

\bibitem{tesla}
Anderson L~I 1994 {\em Nicola Tesla: Lecture Before the New York Academy of
  Sciences - April 6, 1897\/} (Twenty First Century Books)

\bibitem{roger1952double}
Roger B and Georges H 1952 Double heterodyne radio receiver {US Patent
  2,606,285}

\bibitem{protopopov}
Protopopov V~V 2009 {\em Laser Heterodyning\/} (Springer Berlin Heidelberg)

\bibitem{ly2006coherent}
Ly-Gagnon D~S, Tsukamoto S, Katoh K and Kikuchi K 2006 {\em Journal of
  lightwave technology\/} {\bf 24} 12

\bibitem{tanosaki2003vivo}
Tanosaki S, Sasaki Y, Takagi M, Ishikawa A, Inage H, Emori R, Suzuki J, Yuasa
  T, Taniguchi H, Devaraj B {\em et~al.\/} 2003 {\em Optical review\/} {\bf 10}
  447--451

\bibitem{kikuchi2015fundamentals}
Kikuchi K 2015 {\em Journal of Lightwave Technology\/} {\bf 34} 157--179

\bibitem{mecozzi2018coherent}
Mecozzi A and Shtaif M 2018 {\em Optics express\/} {\bf 26} 33970--33981

\bibitem{lu2010distributed}
Lu Y, Zhu T, Chen L and Bao X 2010 {\em Journal of lightwave Technology\/} {\bf
  28} 3243--3249

\bibitem{yuen1983noise}
Yuen H~P and Chan V~W 1983 {\em Optics letters\/} {\bf 8} 177--179

\bibitem{barchielli1990direct}
Barchielli A 1990 {\em Quantum Optics: Journal of the European Optical Society
  Part B\/} {\bf 2} 423

\bibitem{wallentowitz1996unbalanced}
Wallentowitz S and Vogel W 1996 {\em Physical Review A\/} {\bf 53} 4528

\bibitem{grosshans2003quantum}
Grosshans F, Van~Assche G, Wenger J, Brouri R, Cerf N~J and Grangier P 2003
  {\em Nature\/} {\bf 421} 238--241

\bibitem{hirano2003quantum}
Hirano T, Yamanaka H, Ashikaga M, Konishi T and Namiki R 2003 {\em Physical
  Review A\/} {\bf 68} 042331

\bibitem{leverrier2011continuous}
Leverrier A and Grangier P 2011 {\em Physical Review A\/} {\bf 83} 042312

\bibitem{heid2006efficiency}
Heid M and L{\"u}tkenhaus N 2006 {\em Physical Review A\/} {\bf 73} 052316

\bibitem{bradler2018security}
Br{\'a}dler K and Weedbrook C 2018 {\em Physical Review A\/} {\bf 97} 022310

\bibitem{papanastasiou2018quantum}
Papanastasiou P, Lupo C, Weedbrook C and Pirandola S 2018 {\em Physical Review
  A\/} {\bf 98} 012340

\bibitem{ghorai2019asymptotic}
Ghorai S, Grangier P, Diamanti E and Leverrier A 2019 {\em Physical Review X\/}
  {\bf 9} 021059

\bibitem{lin2019asymptotic}
Lin J, Upadhyaya T and L{\"u}tkenhaus N 2019 {\em Physical Review X\/} {\bf 9}
  041064

\bibitem{zhang2019integrated}
Zhang G, Haw J, Cai H, Xu F, Assad S, Fitzsimons J, Zhou X, Zhang Y, Yu S, Wu J
  {\em et~al.\/} 2019 {\em Nature Photonics\/} {\bf 13} 839--842

\bibitem{zhang2020long}
Zhang Y, Chen Z, Pirandola S, Wang X, Zhou C, Chu B, Zhao Y, Xu B, Yu S and Guo
  H 2020 {\em Phys. Rev. Lett.\/} {\bf 125}(1) 010502

\bibitem{merolla1999phase}
Merolla J~M, Mazurenko Y, Goedgebuer J~P, Porte H and Rhodes W~T 1999 {\em
  Optics letters\/} {\bf 24} 104--106

\bibitem{mora2012experimental}
Mora J, Ruiz-Alba A, Amaya W, Mart{\'\i}nez A, Garc{\'\i}a-Mu{\~n}oz V, Calvo D
  and Capmany J 2012 {\em Optics letters\/} {\bf 37} 2031--2033

\bibitem{gleim2016secure}
Gleim A, Egorov V, Nazarov Y~V, Smirnov S, Chistyakov V, Bannik O, Anisimov A,
  Kynev S, Ivanova A, Collins R {\em et~al.\/} 2016 {\em Optics express\/} {\bf
  24} 2619--2633

\bibitem{miroshnichenko2018security}
Miroshnichenko G, Kozubov A, Gaidash A, Gleim A and Horoshko D 2018 {\em Optics
  express\/} {\bf 26} 11292--11308

\bibitem{gaidash2019methods}
Gaidash A, Kozubov A and Miroshnichenko G 2019 {\em JOSA B\/} {\bf 36} B16--B19

\bibitem{chistiakov2019feasibility}
Chistiakov V, Kozubov A, Gaidash A, Gleim A and Miroshnichenko G 2019 {\em
  Optics Express\/} {\bf 27} 36551--36561

\bibitem{kynev2017free}
Kynev S~M, Chistyakov V~V, Smirnov S~V, Volkova K~P, Egorov V~I and Gleim A~V
  2017 {\em Journal of Physics: Conference Series\/} {\bf 917} 052003

\bibitem{samsonov2019continuous}
Samsonov E, Goncharov R, Gaidash A, Kozubov A, Egorov V and Gleim A 2020 {\em
  Scientific Reports\/} {\bf 10} 10034

\bibitem{fang2014multichannel}
Fang J, Huang P and Zeng G 2014 {\em Physical Review A\/} {\bf 89} 022315

\bibitem{gyongyosi2019subcarrier}
Gyongyosi L and Imre S 2019 {\em Journal of Statistical Physics\/} {\bf 177}
  960--983

\bibitem{wang2019optical}
Wang Y, Mao Y, Huang W, Huang D and Guo Y 2019 {\em Optics express\/} {\bf 27}
  25314--25329

\bibitem{haykin2008communication}
Haykin S 2008 {\em Communication systems\/} (John Wiley \& Sons)

\bibitem{miroshnichenko2017algebraic}
Miroshnichenko G~P, Kiselev A~D, Trifanov A~I and Gleim A~V 2017 {\em JOSA B\/}
  {\bf 34} 1177--1190

\bibitem{capmany2010quantum}
Capmany J and Fern{\'a}ndez-Pousa C~R 2010 {\em JOSA B\/} {\bf 27} A119--A129

\bibitem{kumar2008evolution}
Kumar P and Prabhakar A 2008 {\em IEEE journal of quantum electronics\/} {\bf
  45} 149--156

\bibitem{yariv1984optical}
Yariv A and Yeh P 1984 {\em Optical waves in crystals\/} vol~5 (Wiley New York)

\bibitem{gonzalez2010optical}
Gonzalez N~G, Zibar D, Yu X and Monroy I~T 2010 Optical phase-modulated
  radio-over-fiber links with k-means algorithm for digital demodulation of
  8psk subcarrier multiplexed signals {\em Optical Fiber Communication
  Conference\/} (Optical Society of America) p OML3

\bibitem{gasulla2012phase}
Gasulla I and Capmany J 2012 {\em Optics express\/} {\bf 20} 11710--11717

\bibitem{zhang2008photonic}
Zhang Y, Xu K, Zhu R, Li J, Wu J, Hong X and Lin J 2008 {\em Optics letters\/}
  {\bf 33} 2332--2334

\bibitem{hirano2017implementation}
Hirano T, Ichikawa T, Matsubara T, Ono M, Oguri Y, Namiki R, Kasai K, Matsumoto
  R and Tsurumaru T 2017 {\em Quantum Science and Technology\/} {\bf 2} 024010

\bibitem{leverrier2009theoretical}
Leverrier A 2009 {\em Theoretical study of continuous-variable quantum key
  distribution\/} Ph.D. thesis T{\'e}l{\'e}com ParisTech

\bibitem{yoshida200864}
Yoshida M, Goto H, Kasai K and Nakazawa M 2008 {\em Optics Express\/} {\bf 16}
  829--840

\bibitem{hongo20071}
Hongo J, Kasai K, Yoshida M and Nakazawa M 2007 {\em IEEE Photonics Technology
  Letters\/} {\bf 19} 638--640

\bibitem{gebbie1967heterodyne}
Gebbie H, Stone N, Putley E and Shaw N 1967 {\em Nature\/} {\bf 214} 165--166

\bibitem{maznev1998optical}
Maznev A, Nelson K and Rogers J~A 1998 {\em Optics letters\/} {\bf 23}
  1319--1321

\bibitem{fried1967optical}
Fried D~L 1967 {\em Proceedings of the IEEE\/} {\bf 55} 57--77

\bibitem{delange1968optical}
DeLange O 1968 {\em IEEE spectrum\/} {\bf 5} 77--85

\bibitem{bylina2017comparative}
Bylina M, Glagolev S and Diubov A 2017 {\em Proceedings of {T}elecommunication
  {U}niversities\/} {\bf 3} 21--28

\bibitem{wang2015iq}
Wang C, Qu Y and Tang Y~P~T 2015 {\em Infrared Physics \& Technology\/} {\bf
  72} 191--194

\bibitem{bohme1999heterodyne}
Bohme R and Eichin M 1999 Heterodyne receiver with synchronous demodulation for
  receiving time signals {US Patent 5,930,697}

\bibitem{chen2010optical}
Chen Y~K, Koc U~V and Leven A 2010 Optical heterodyne receiver and method of
  extracting data from a phase-modulated input optical signal {US Patent
  7,650,084}

\bibitem{chi2011balanced}
Chi Y~M, Qi B, Zhu W, Qian L, Lo H~K, Youn S~H, Lvovsky A and Tian L 2011 {\em
  New Journal of Physics\/} {\bf 13} 013003

\bibitem{tang2020performance}
Tang X, Kumar R, Ren S, Wonfor A, Penty R and White I 2020 {\em Optics
  Communications\/}  126034

\end{thebibliography}

\end{document}